# A Sequential Detection and Tracking of Very Low SNR Objects

Reza Rezaie[1]

**Abstract**- A sequential detection and tracking (SDT) approach is proposed for detection and tracking of very low signal-to-noise (SNR) objects. The proposed approach is compared with two existing particle filter track-before-track (TBD) methods. It is shown that the former outperforms the latter. A conventional detection and tracking (CDT) approach, based on one-data-frame thresholding, is considered as a benchmark for comparison. Simulations demonstrate the performance.

I. INTRODUCTION

Detection and tracking of a low SNR object is challenging. The SNR is not high enough to detect the object based on thresholding the signal intensity at each data frame. Accumulation of signal from an object over frames/time increases the SNR and makes the object detectable. Considering a sequence of data frames as input, in conventional detection and tracking (CDT) methods, detection is based on thresholding the signal intensity at each frame and then once an object is detected its position is estimated with a higher accuracy using a tracker. However, if the SNR of the object is not high enough to be detected based on one-frame thresholding, several consecutive frames of measurements can be used to accumulate signal over frames/time to enhance the SNR and detect the object with a high probability of detection. Since some tracking of the object is needed for signal accumulation before its detection, such approaches are sometimes called track-before-detect (TBD). Several TBD approaches have been presented in the literature using dynamic programming, Hough transform, maximum likelihood, particle filters, etc. [1]-[18]. In [7], an intensity marginalized likelihood ratio along with the finite set statistics approach were used to sequentially detect and track low SNR objects. Also, [8] used a sequential detection method to enhance a nearest neighbor tracker and improve data to track correlation for low SNR environments. The approaches of [7] and [8] are different from the one proposed in this paper. For example, in our approach there is neither finite set statistics nor sequential detection-based enhancement of a nearest neighbor tracker. The proposed approach is briefly explained in the next paragraph, and its details are provided in Section IV. In particle filter TBD approaches the whole detection and tracking problem is handled as a density estimation problem in the particle filter framework. Particle filter TBDs have demonstrated their effectiveness [9]-[18].

---

[1] rezarezaie01@gmail.com

This paper presents a detection and tracking approach for very low SNR objects where signals are accumulated over frames/time in a sequential framework to enhance the SNR, where all possible behaviors/motions of the object are considered. Once an object is detected, its information is sent to a probabilistic data association particle filter tracker for a more accurate estimation of its state. It is shown that the proposed approach can detect and track very low SNR objects and outperforms the existing particle filter TBDs.

In the following, sensor and motion models are reviewed in Section II. A conventional one-frame-thresholding detection and tracking approach is briefly reviewed in section III. The proposed SDT is presented in section IV. Then, two existing particle filter TBD approaches are briefly reviewed in section V. In section VI, the SDT, two particle filter TBD, and the conventional detection and tracking (CDT) approaches are compared. For simplicity of our formulation, it is assumed that the object is a point and does not contribute to more than one frame cell. Also, it is assumed that there is at most one object in the region of interest.

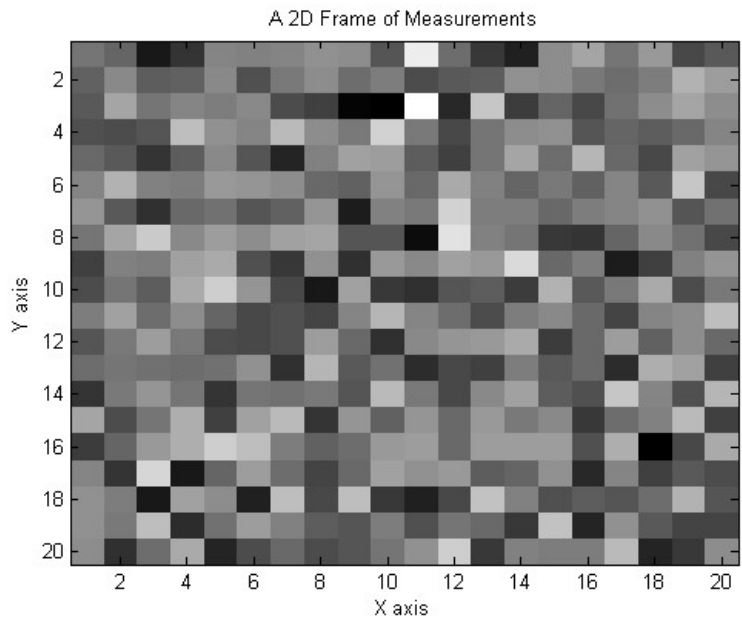

Fig 1: A data frame with 20*20 pixels/cells

## II. OBJECT MOTION AND SENSOR MODELING

**Object Motion Model**

Object motion model is as follows,

$$x_k = F x_{k-1} + v_k \quad F = \begin{bmatrix} 1 & T & 0 & 0 \\ 0 & 1 & 0 & 0 \\ 0 & 0 & 1 & T \\ 0 & 0 & 0 & 1 \end{bmatrix} \quad (1)$$

$$x_k = [p_{x,k}, \dot{p}_{x,k}, p_{y,k}, \dot{p}_{y,k}]'$$

in which $x_k$ is the state vector including position and velocity in 2D space, $v_k$ is the process noise, and F is the transition matrix.

**Sensor Model**

The received intensity in each cell is modeled as follows [21],

$$z_k^{(i,j)} = h_k^{(i,j)}(x_k) + w_k^{(i,j)} \quad (2)$$

in which

$$h_k^{(i,j)}(x_k) = \begin{cases} \dfrac{\Delta_x \Delta_y I_k}{2\pi \Sigma^2} \exp\left\{ -\dfrac{(i\Delta_x - p_{x,k})^2 + (j\Delta_y - p_{y,k})^2}{2\Sigma^2} \right\} & \text{if object present in cell (i,j) (i.e. } H_1) \\ 0 & \text{otherwise (i.e. } H_0) \end{cases}$$

$\Delta_x \times \Delta_y$: Cell dimension $\quad (i\Delta_x, j\Delta_y), i = 1, \ldots, n, j = 1, \ldots, m$

$w_k^{(i,j)}: N(0, \sigma^2) \quad$ Measurement noise in cell (i,j) at time k

$I_k$: Object Signal Intensity

$\Sigma$ : Sensor bluring

And the measurement of each image frame is as follows,

$$Z_k = \left\{ z_k^{(i,j)} : i = 1, \ldots, n, j = 1, \ldots, m \right\}$$

All frames since the beginning to time k is denoted as

$$Z^k = \{Z_i, i = 1, \ldots, k\}$$

For simplicity we consider no sensor blurring, i.e.,

$$h_k^{(i,j)}(x_k) = \begin{cases} I_k & \text{if object present in cell (i,j)} \\ 0 & \text{otherwise} \end{cases} \quad (3)$$

## III. COVENTIONAL DETECTION AND TRACKING (CDT)

CDT is reviewed in this section as a benchmark.

The likelihood ratio test for each cell at frame $k$ is

$$\begin{cases} L(z_k^{i,j}) > th & : H_1 \\ L(z_k^{i,j}) < th & : H_0 \end{cases} \quad (4)$$

where

$$L(z_k^{i,j}) = \frac{f(z_k^{i,j}|H_1)}{f(z_k^{i,j}|H_0)}, \quad i = 1,\ldots,n, j = 1,\ldots,m$$

Based on the sensor model we have

$$\frac{f(z_k^{i,j}|H_1)}{f(z_k^{i,j}|H_0)} = exp(-\frac{I_k(I_k - 2z_k^{i,j})}{2\sigma^2})$$

And the final test statistics is

$$\begin{cases} z_k^{i,j} > th' & : H_1 \\ z_k^{i,j} < th' & : H_0 \end{cases}$$

Then, if an object is detected in a cell, the cell information is sent to a tracker to have a more accurate estimate of the object's state. The tracker is explained in the next section.

## IV. PROPOSED SEQUENTIAL DETECTION AND TRACKING (SDT)
### A. Sequential Detection (SD)

In this approach measurements are accumulated over frames/time using an SD framework to enhance the SNR of an object and detect it. The SD is based on the following likelihood ratio which is a function of measurements in $M$ consecutive frames.

$$\begin{aligned}
L_M < th_1 & \qquad : H_0 \\
L_M > th_2 & \qquad : H_1 \\
th_1 < L_M < th_2 & \qquad \text{no decision (wait for next frame)}
\end{aligned} \quad (5)$$

where

$$L_M = \prod_{n=1}^{M} \frac{f(z_n^{i_n,j_n}|H_1)}{f(z_n^{i_n,j_n}|H_0)}$$

$M$ is the total number of sequential iterations, i.e., number of frames used for measurement accumulation. The subscript, $n$, is the iteration counter (i.e. the $n$th frame used for measurement accumulation) in the SD framework, and $(i_n, j_n)$ is the cell (the measurement $z_n^{i_n,j_n}$ comes from) at the $n$th iteration. For simplicity we drop the time/frame subscript.

**Remark:** The subscript $k$ in Section III is frame/time whereas the subscript $n$ in Section IV is the sequential iteration.

Measurements used in $L_M$ are

$$z_1^{i_1,j_1}, z_2^{i_2,j_2}, z_3^{i_3,j_3}, \ldots, z_M^{i_M,j_M}$$

After some manipulation of the likelihood ratio we have,

$$\begin{cases}
T_M < th_1' & \qquad : H_0 \\
T_M > th_2' & \qquad : H_1 \\
th_1' < T_M < th_2' & \qquad \text{no decision (wait for next frame)}
\end{cases} \quad (6)$$

where

$$T_M = \sum_{n=1}^{M} z_n^{i_n,j_n} \quad (7)$$

The idea of this SD framework is that by accumulating enough measurements from a low SNR object in a cell the SNR enhances, and the object can be detected with a high probability. However, the problem is that it is possible that before having enough measurements from an object in a cell to have a high enough SNR, the object may move to a different cell. In other words, it is not necessarily possible to accumulate enough measurements from the object while it is staying in one cell. To address this issue, it is assumed that the object may move at most one cell between two consecutive frames and the likelihood ratio is calculated for all such possibilities in all directions around a cell. That is how $i_n$ and $j_n$ in (4) are determined. The computational complexity of this approach grows quickly. So, depending on the available computational resources, a maximum value is considered for the number of iterations $M$ in the SD. This maximum is denoted by $S_m$. If the SD framework reaches its maximum iteration while the likelihood ratio is still between the two thresholds, a truncated SD is used [19], and the decision is made based on the following test statistics

$$\begin{cases} T_{s_m} > \tau & : H_1 \\ T_{s_m} < \tau & : H_0 \end{cases} \qquad (8)$$

As mentioned above, it is assumed that an object may move at most one cell between two consecutive frames. This assumption can be extended to more than one cell movement at the cost of more computational complexity. Depending on available computational resources, one can divide the region of interest into some sub-regions and use parallel processing.

To illustrate how effective the proposed SD approach is, assume the object signal intensity and the measurement noise variance are constant over frames, i.e., $I_k = I$ and $\sigma_k^2 = \sigma^2$. By each frame signal accumulation, both the object signal intensity and the noise variance double. Therefore, the signal to noise ratio doubles, i.e., $(2 \times I)^2 / (2 \times \sigma^2) = 2 \times (I^2 / \sigma^2)$.

A summary of the above SD framework:

- At the first measurement frame, select the cells that may contain an object (cells with signal intensity higher than a very low threshold) and calculate $T_M$ as in (7), where $M = 1$. For $M < s_m$, based on (6) either a decision is made or there is no decision and wait for the next frame. In case $M = s_m$, based on (8) a decision is made. Once an object is detected, its cell information is sent to the tracker.
- At the next frame, consider the selected cells from the previous frame and select one-cell vicinity of them (including themselves) in all directions and calculate $T_2$ for each one. Also, among those cells that their measurements have not been used, select the ones with signal intensity higher than a very low threshold and calculate $T_1$ for each one. For $M < s_m$, based on (6) either a decision is made or there is no decision and wait for the next frame. In case $M = s_m$, based on (8) a decision is made. Once an object is detected, its cell information is sent to the tracker.
- At the next frame, consider the selected cells from the previous frame and select one-cell vicinity of them (including themselves) in all directions and calculate $T_3$ or $T_2$ (depending on whether it is the 3$^{rd}$ or the 2$^{nd}$ iteration in the SD framework) for each one. Also, among those cells that their measurements have not been used, select those with signal intensity higher than a very low threshold and calculate $T_1$ for each one. For $M < s_m$, based on (6) either a decision is made or there is no decision and wait for the next frame. In case $M = s_m$, based on (8) a decision is made. Once an object is detected, its cell information is sent to the tracker.
- At the next frame, consider the selected cells from the previous frame and select one-cell vicinity of them (including themselves) in all directions and calculate $T_4$, $T_3$, or $T_2$ (depending on whether it is the 4$^{th}$, the 3$^{rd}$, or the 2$^{nd}$ iteration in the SD framework) for each one. Also, among those cells that their measurements have not been used, select those with signal intensity higher than a very low threshold and calculate $T_1$ for each one. For $M < s_m$, based on (6) either a decision is made or there is no decision and wait for the next frame. In case $M = s_m$, based on (8) a decision is made. Once an object is detected, its cell information is sent to the tracker.
- Continue the above approach for the next frames.

See [19] for general sequential analysis approaches.

## B. Tracking

The cell information of detected objects is sent to a probabilistic data association multiple model particle filter (PDA-MMPF) to provide a more accurate estimate of its position and velocity. In this approach data association is based on a probabilistic data association (PDA) method [20] and the filtering is based on a multiple model particle filter (MMPF) [21]. More details of PDA and MMPF can be found in [20] and [21], respectively. Also, see [21]-[25] for details about particle filters.

## V. TRACK-BEFORE-DETECT PARTICLE FILTERS

Two existing particle filter TBD approaches [9]-[11] are reviewed and later compared with the proposed SDT approach in Section VI. Both particle filter TBD approaches have the following assumptions:

- A variable is defined as object existence $E_k$ and is modeled as a two-state Markov chain, where $E_k = 1$ means object existence and $E_k = 0$ means no object.
- Transition probability of birth and death is
$$\Pi_E = \begin{bmatrix} 1 - P_b & P_b \\ P_d & 1 - P_d \end{bmatrix}$$
where
$P_b = Pr\{E_k = 1 | E_{k-1} = 0\}$
$P_d = Pr\{E_k = 0 | E_{k-1} = 1\}$
- The initial object existence probability (at time k =1) is assumed known (design parameters).

Based on the above assumptions, the ideas behind the two particle filter TBD approaches are briefly reviewed. Details are skipped. They can be found in [9]-[11].

### A. First Particle Filter TBD Approach (TBD1)

This approach is based on a recursive Bayesian estimation of the joint density of the state and the existence [9], i.e.,

$$p(x_{k-1}, E_{k-1}|Z^{k-1}) \to p(x_k, E_k|Z^k)$$

where $Z^k = \{Z_i, i = 1, \ldots, k\}$. Then, using the joint density $p(x_k, E_k = 1|Z^k)$ the posterior marginal density of the existence variable is used as a test statistic and is compared with a threshold. If the posterior existence density is higher than a threshold an object is declared in the cell, otherwise it is claimed that there is no object, i.e.,

$$\begin{cases} p(E_k = 1|Z^k) > th & : H_1 \\ p(E_k = 1|Z^k) < th & : H_0 \end{cases}$$

If an object is declared in a cell, its state estimate is obtained based on the posterior density

$$p(x_k|E_k = 1, Z^k)$$

TBD1 does not have any logic to handle maneuvering objects. We extend TBD1 to a multiple model TBD1 (MMTBD1) where switching between different motion dynamic models is modeled by a Markov chain variable $r_k \in \{1, \ldots, m\}$. This Markov chain is handled in the recursive Bayesian estimation framework along with the state and the existence variable in the following joint posterior density [26]

$$p(x_k, r_k, E_k|Z^{k-1})$$

We skip the details of the recursive Bayesian propagation of this posterior density. A particle filter implementation of MMTBD1 is used in simulations for comparison.

B. **Second Particle Filter TBD Approach (TBD2)**

This approach estimates the joint density of the state and the presence probability [11]. Then if based on the presence probability estimation an object is declared, the state vector is estimated. Joint density of the presence probability and the state vector using all the observations since the beginning to the current time is as follows

$$p(x_k, E_k|Z^k) = p(x_k|E_k, Z^k)p(E_k|Z^k) \qquad (9)$$

Since density of the state is calculated only if an object is present, and since sum of the presence and absence probabilities is 1, it is enough to just compute the following terms

$$p(x_k|E_k = 1, Z^k), \quad p(E_k = 1|Z^k)$$

TBD2 does not have any logic to handle maneuvering objects [11]. We extend TBD2 to a multiple model TBD2 (MMTBD2) where switching between different motion dynamic models is modeled by a Markov chain variable $r_k \in \{1, \ldots, m\}$. This Markov chain is handled in the recursive Bayesian estimation framework along with the state and the existence variable in the following joint posterior density [26]

$$p(x_k, r_k, E_k | Z^k) = p(x_k, r_k | E_k, Z^k) p(E_k | Z^k)$$

Similar to TBD2, it is enough to have a recursive calculation of the following two terms

$$p(x_k, r_k | E_k = 1, Z^k), \quad p(E_k = 1 | Z^k)$$

We skip the details of the recursive Bayesian propagation of these posterior densities. A particle filter implementation of MMTBD2 is used in simulations for comparison.

## VI. SIMULATIONS

We first compared MMTBD1 and MMTBD2, where MMTBD2 outperforms MMTBD1. Therefore, we only consider MMTBD2 in this section to compare with the CDT and the proposed SDT approaches.

Several examples are presented. In the first example MMTBD2 and TBD2 are compared to show how our multiple model extension of TBD2 improved its performance. In the second example MMTBD2 and CDT are compared. In the third example MMTBD2 and SDT are compared. In the fourth example the two approaches are compared in a high speed and high maneuver scenario. In the fifth example SDT is evaluated in very low SNR cases to show it can even handle such objects. The CDT is a special case of SDT where only one frame is considered. So, their performances are compared to evaluate the latter. A scenario with 40 frames of measurements is presented in each example. Also, for illustration, different SNRs are considered.

**Example 1:** MMTBD2 and TBD2 are compared to demonstrate the impact of multiple modeling for handling maneuvers.

- There exists an object from frame 10 to frame 35.
- Two cases with two different SNRs (6dB and 12dB) are considered.
- Probability of detection during the whole 40 frames and RMSE of position and velocity estimates (when an object is detected) are plotted.
- MMTBD2 outperforms TBD2.

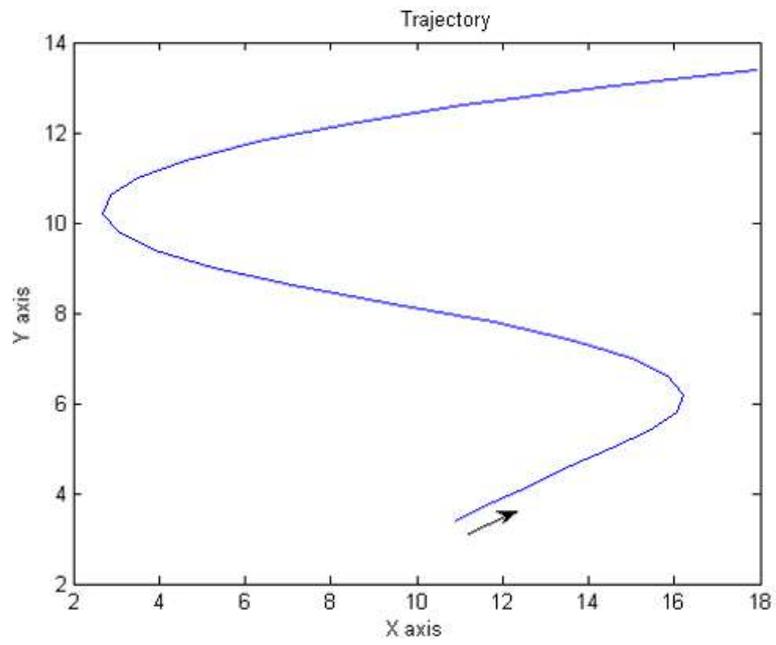

Fig 2: Trajectory

**Case I: SNR= 6dB**

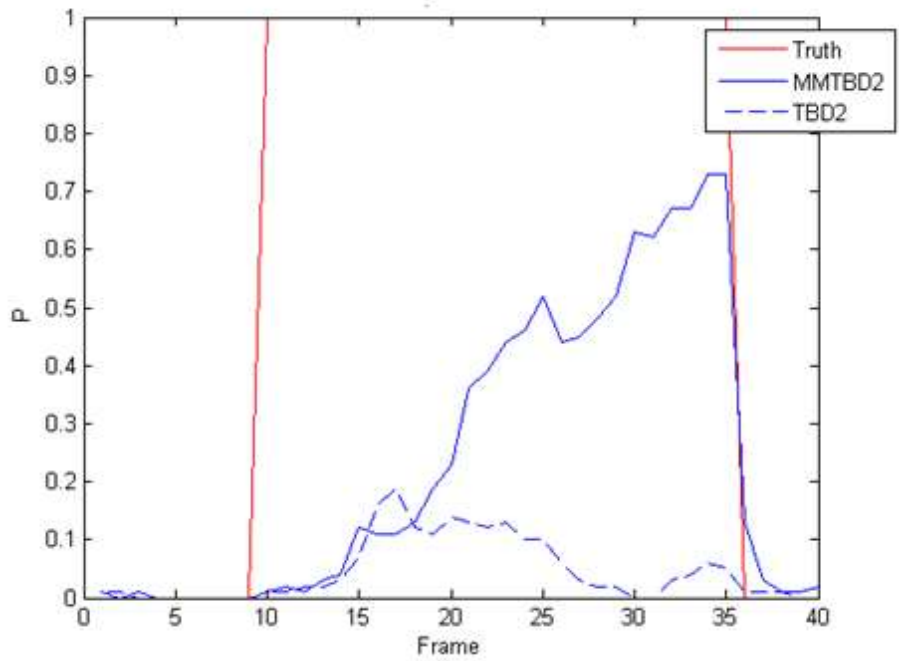

Fig 3: Detection probability

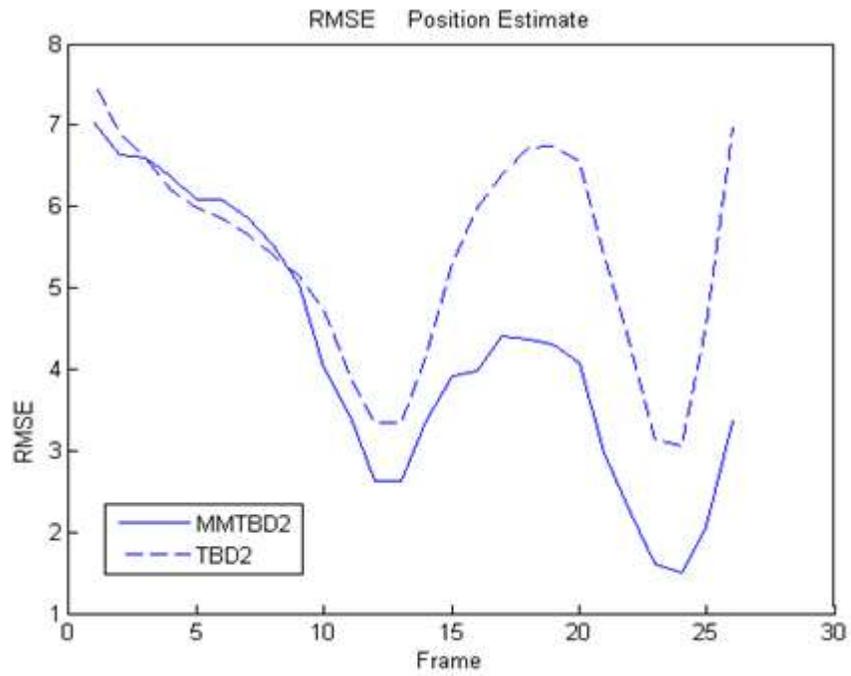

Fig 4: RMSE position estimate

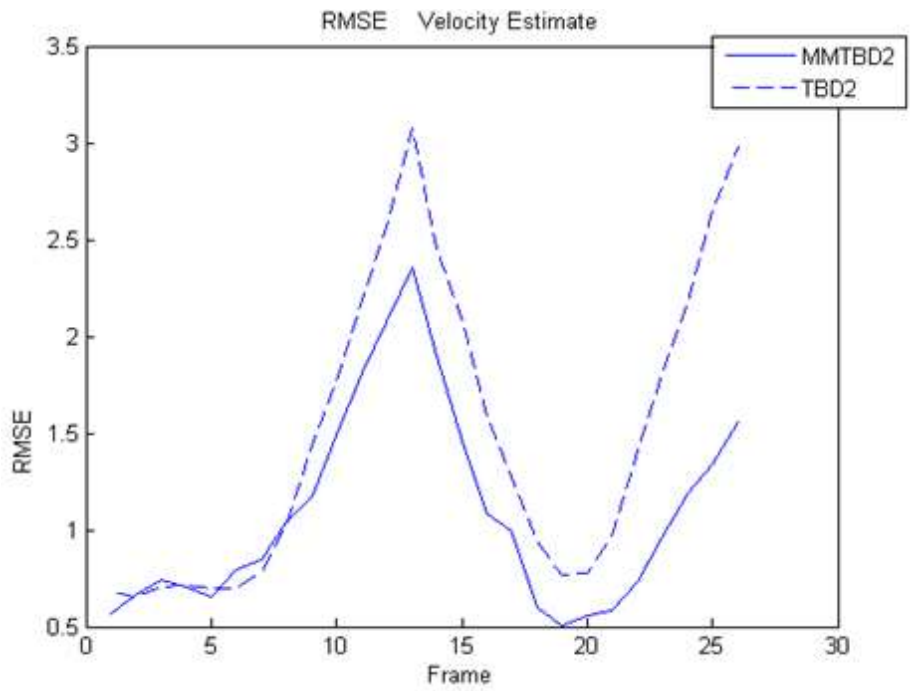

Fig 5: RMSE velocity estimate

**Case II: SNR=12dB**

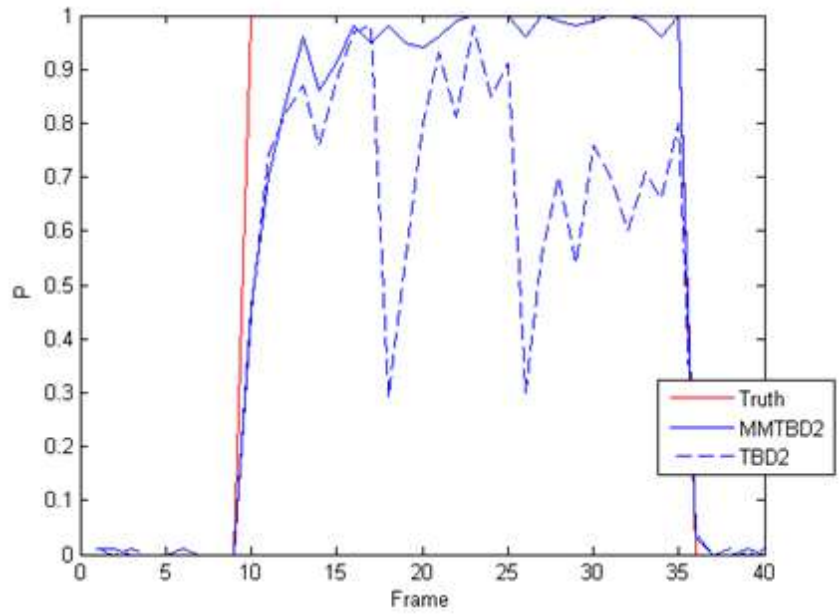

Fig 6: detection probability

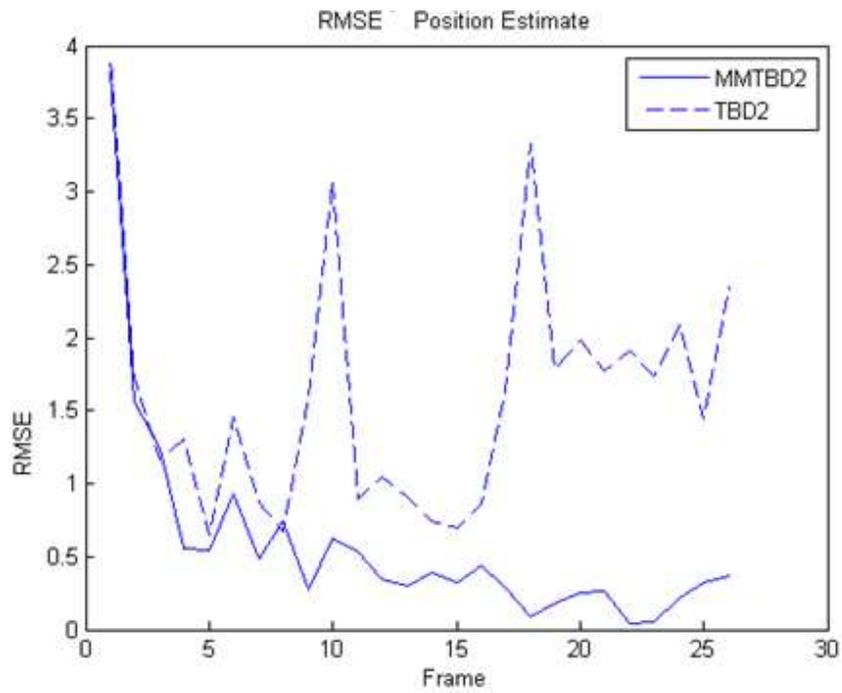

Fig 7: RMSE of position estimate

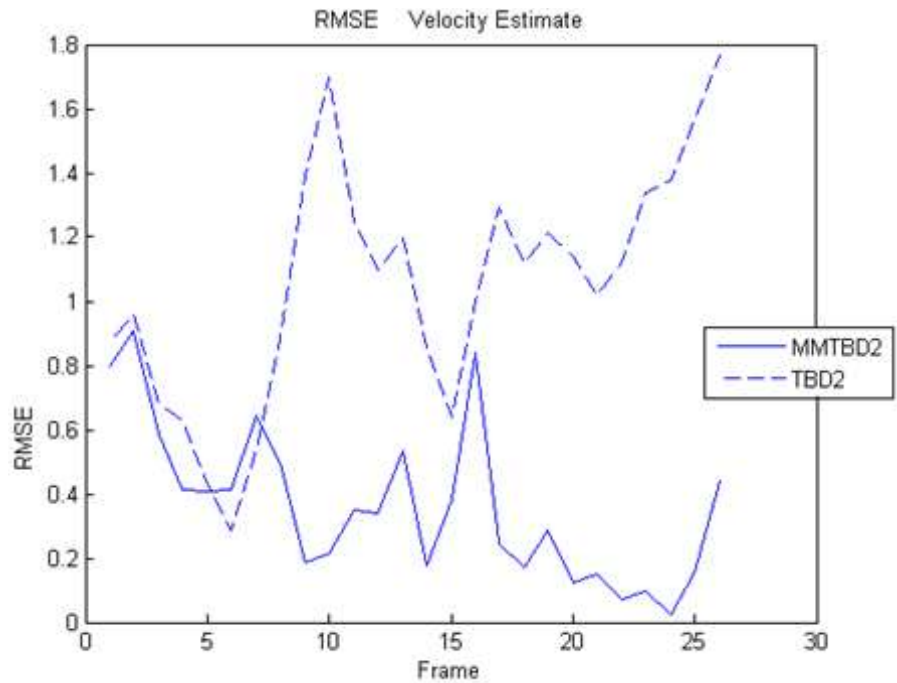

Fig 8: RMSE of velocity estimate

**Example 2:** MMTBD2 and CDT are compared.

- There exists an object from frame 10 to frame 35.
- SNR=5dB.
- Probability of detection during the whole 40 frames and RMSE of position and velocity estimates (when an object is detected) are plotted.
- MMTBD2 outperforms CDT in detection, but not in tracking.

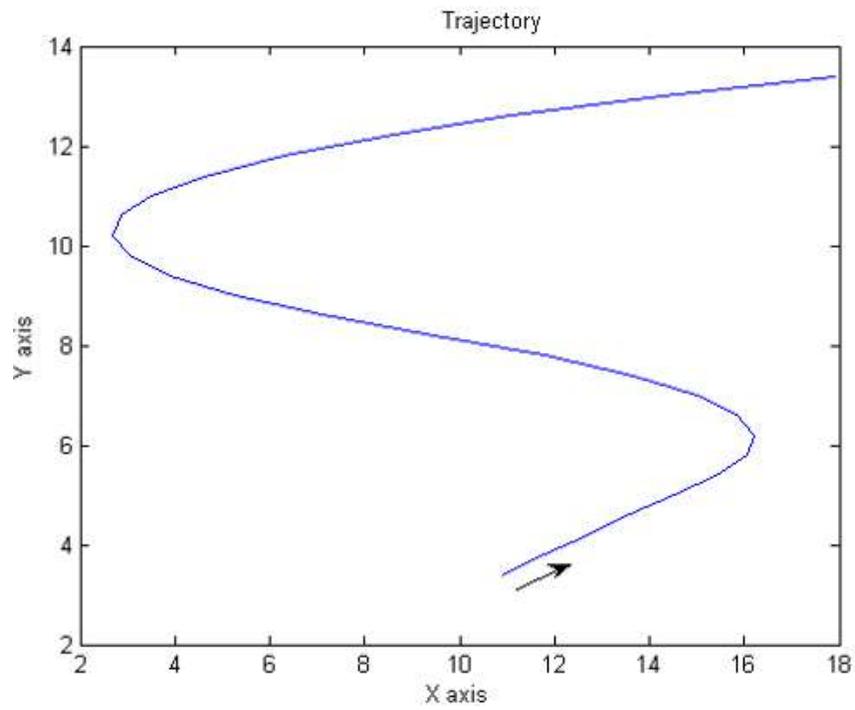

Fig 9: Trajectory

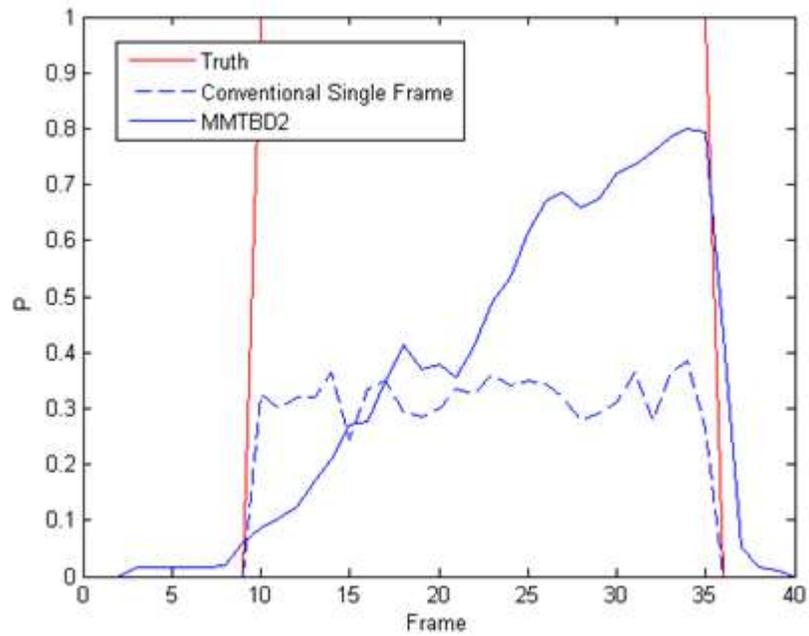

Fig 10: Detection probability

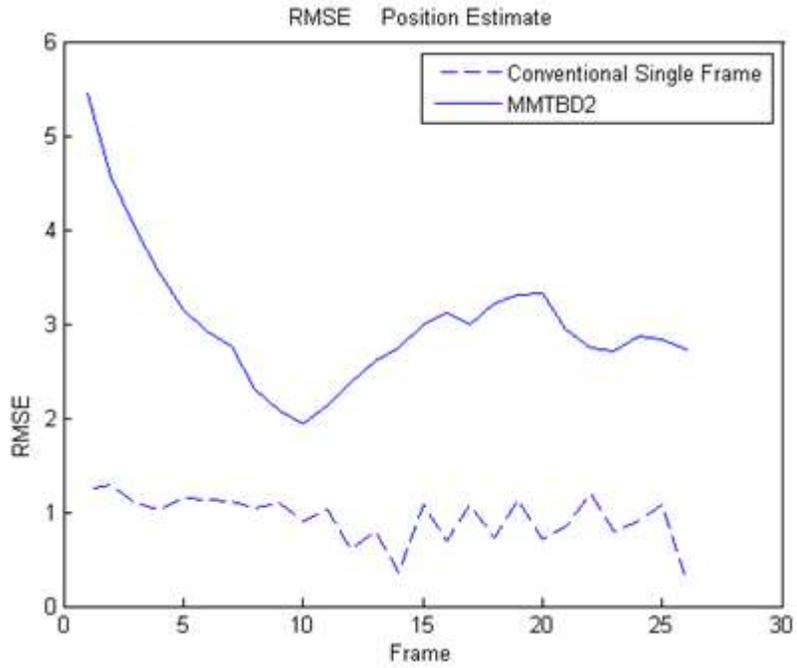

Fig 11: RMSE of position estimate

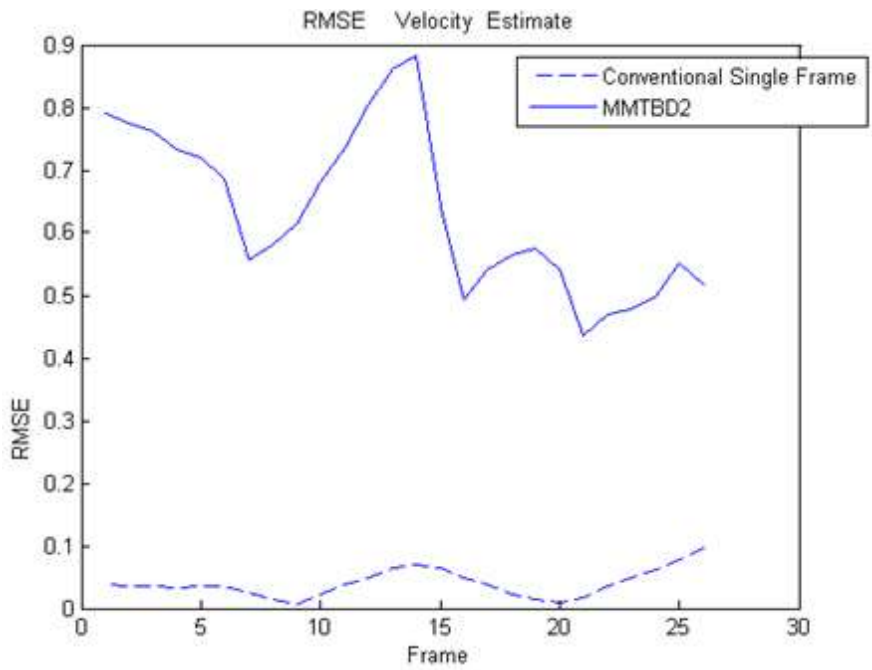

Fig 12: RMSE of velocity estimate

**Example 3:** SDT and CDT are compared.

- There exists an object from frame 5 to frame 35.
- SNRs 3dB and 6dB.
- Probability of detection during the whole 40 frames and RMSE of position and velocity estimates (when an object is detected) are plotted.
- SDT outperforms CDT.

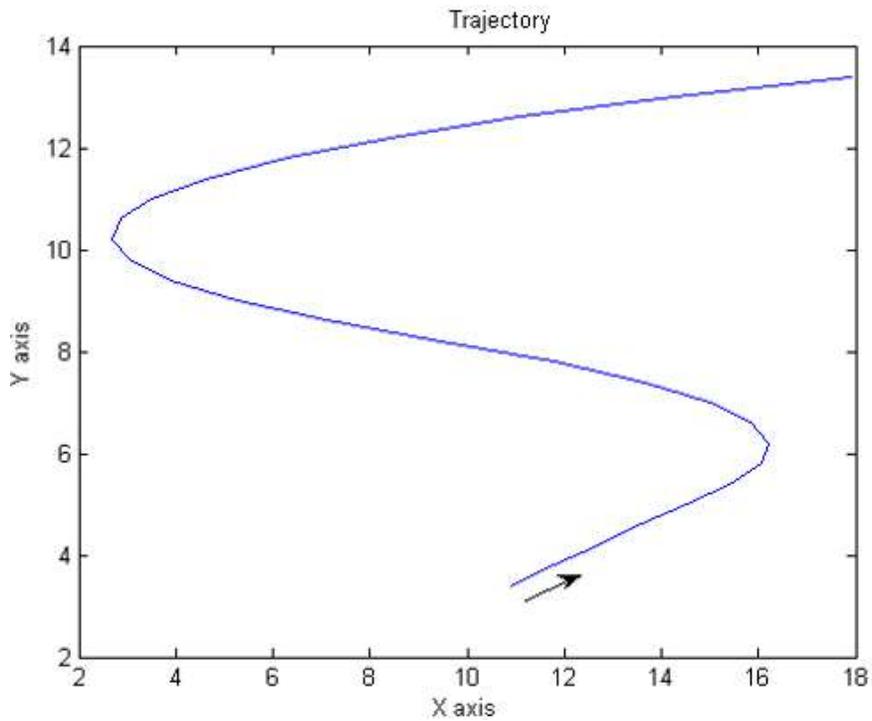

Fig 13: Trajectory

**Case I: SNR: 3dB**

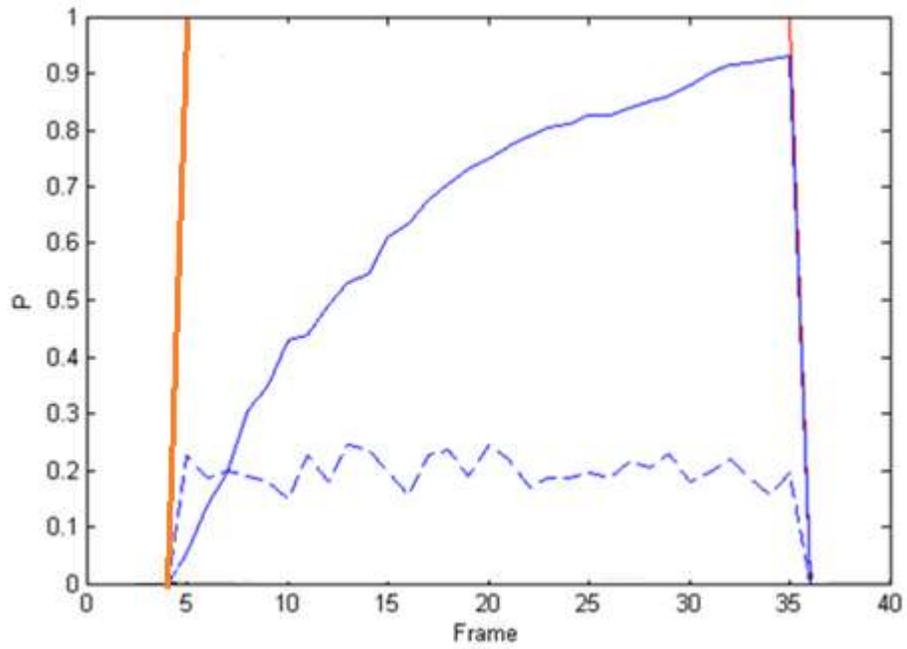

Fig 14: Detection probability (dash line: CDT, solid line: SDT)

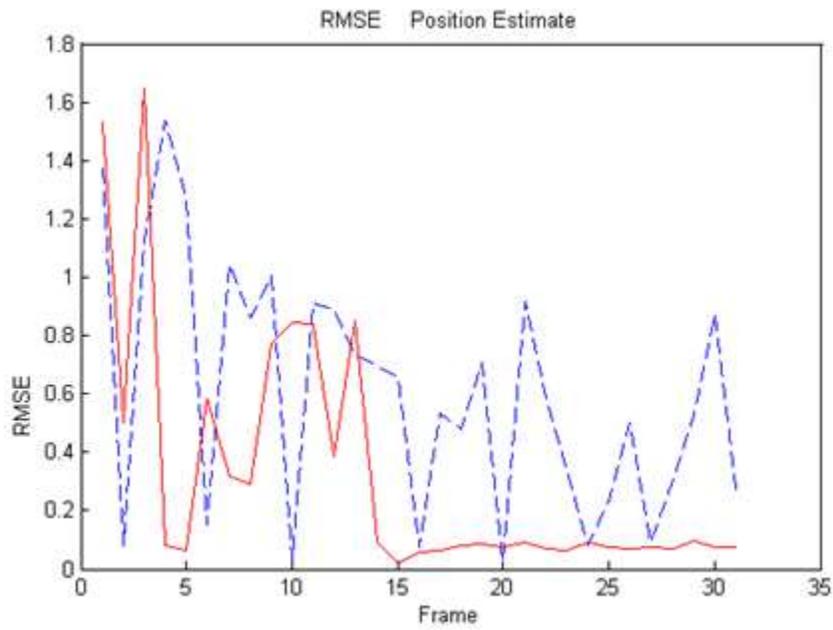

Fig 15: RMSE of position estimate (dash line: CDT, solid line SDT)

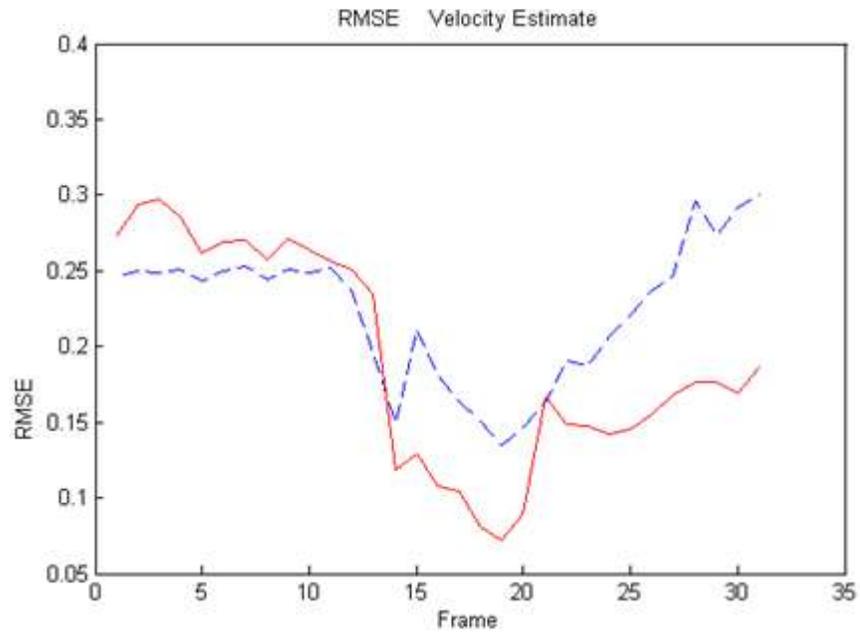

Fig 16: RMSE of velocity estimate (dash line: CDT, solid line: SDT)

**Case II: SNR: 6dB**

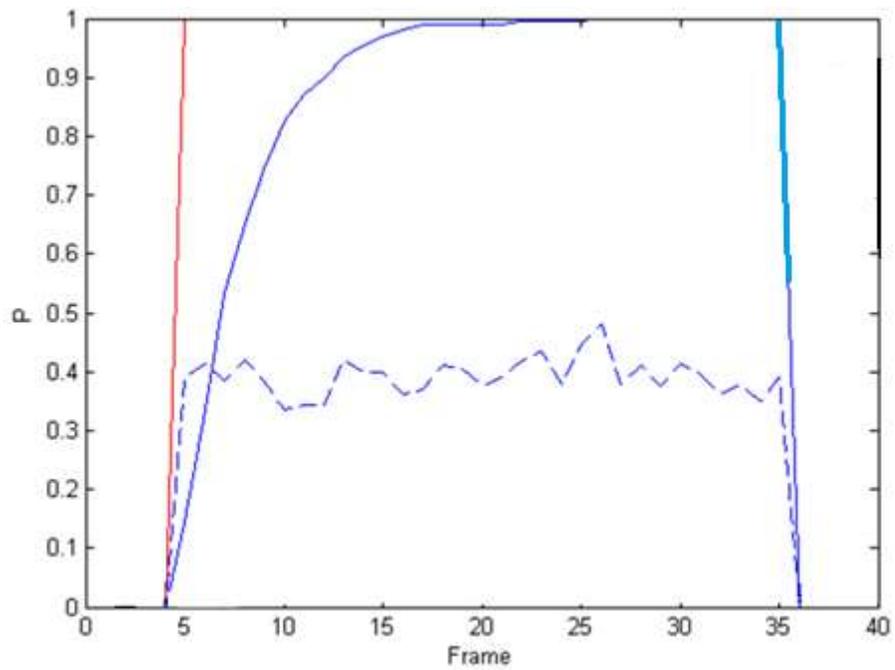

Fig 17: Detection probability (dash line: CDT, solid line: SDT)

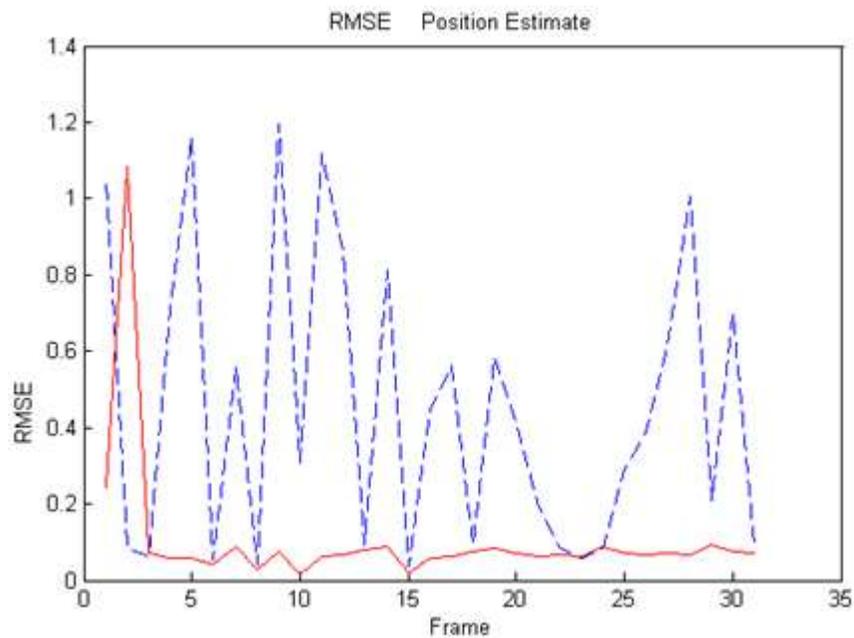

Fig 18: RMSE of position estimate (dash line: CDT, solid line: SDT)

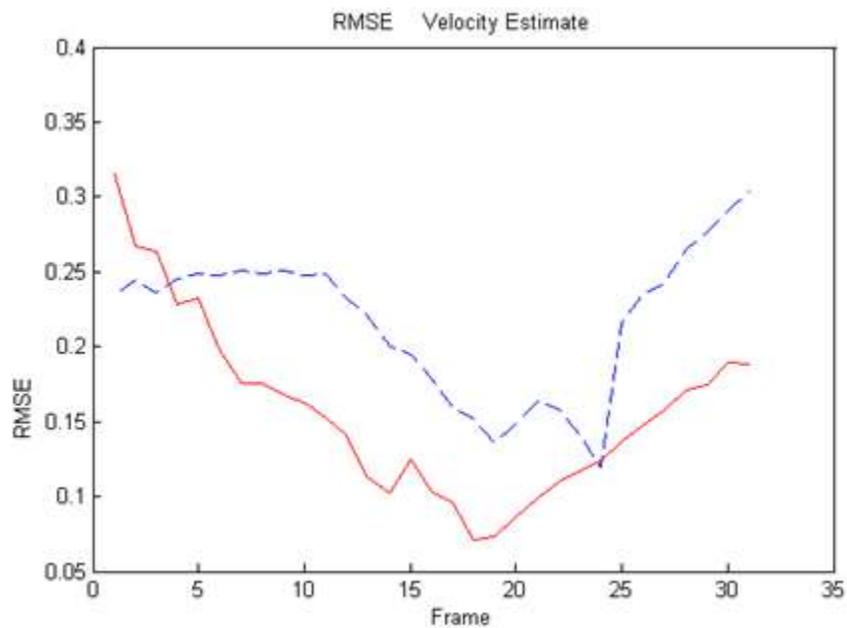

Fig 19: RMSE of velocity estimate (dash line: CDT, solid line: SDT)

**Example 4:** SDT and MMTBD2 are compared.

- There exists an object from frame 5 to frame 35.

- SNRs 3dB, 5dB, and 9dB.
- Probability of detection during the whole 40 frames and RMSE of position and velocity estimates (when an object is detected) are plotted.
- $s_m = 3$.
- SDT outperforms MMTBD2.

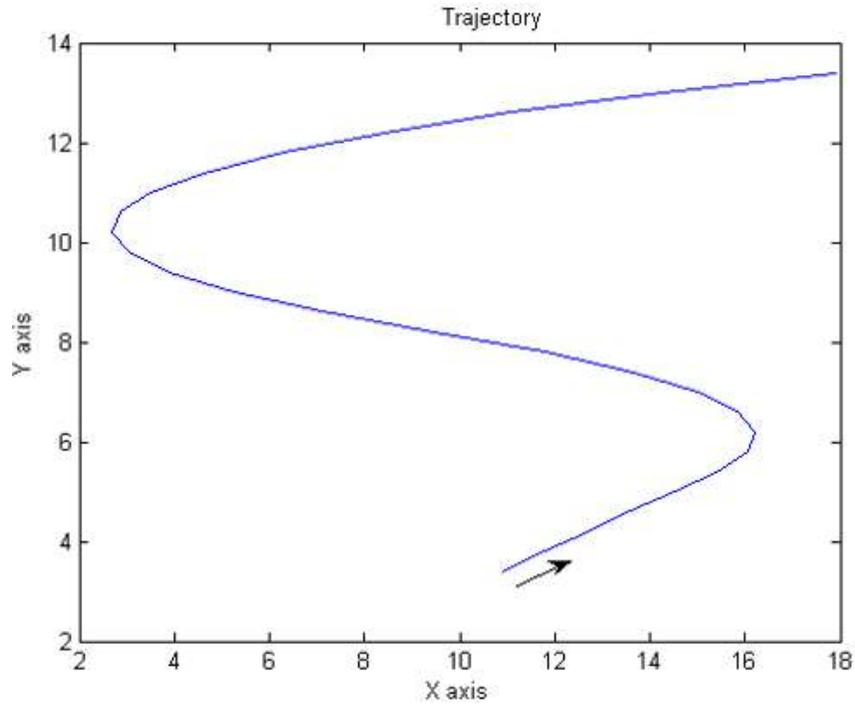

Fig 20: Trajectory

**Case I: SNR=3dB**

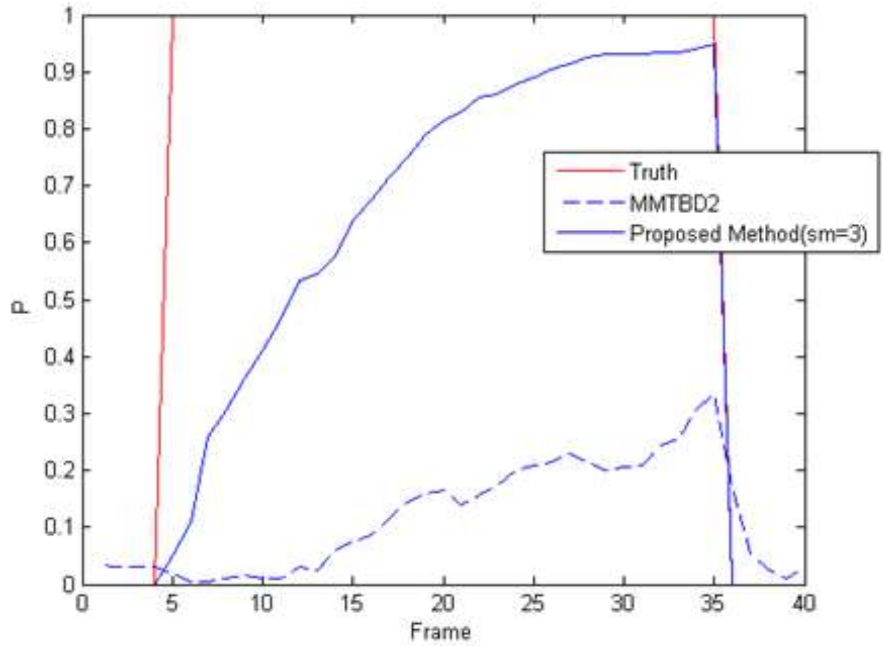

Fig 21: Detection probability (dash line: MMTBD2, solid line: SDT)

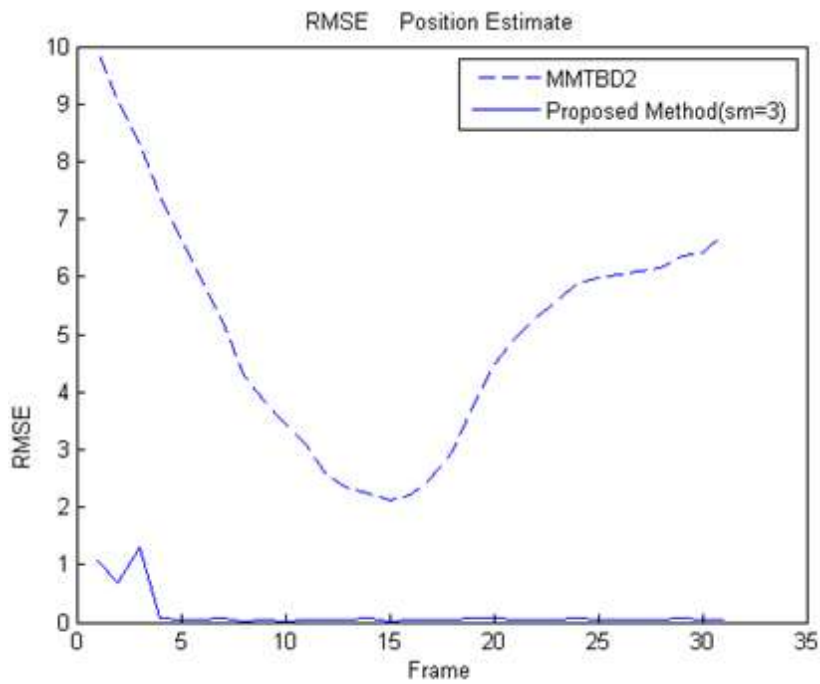

Fig 22: RMSE of position estimate (dash line: MMTBD2, solid line: SDT)

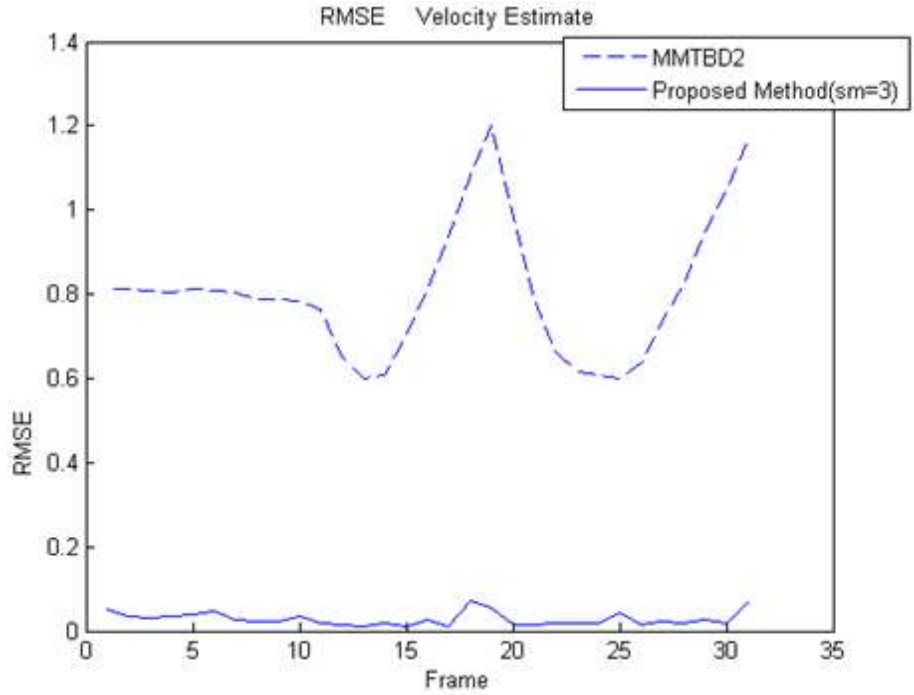

Fig 23: RMSE of velocity estimate (dash line: MMTBD2, solid line: SDT)

**Case II: SNR=5dB**

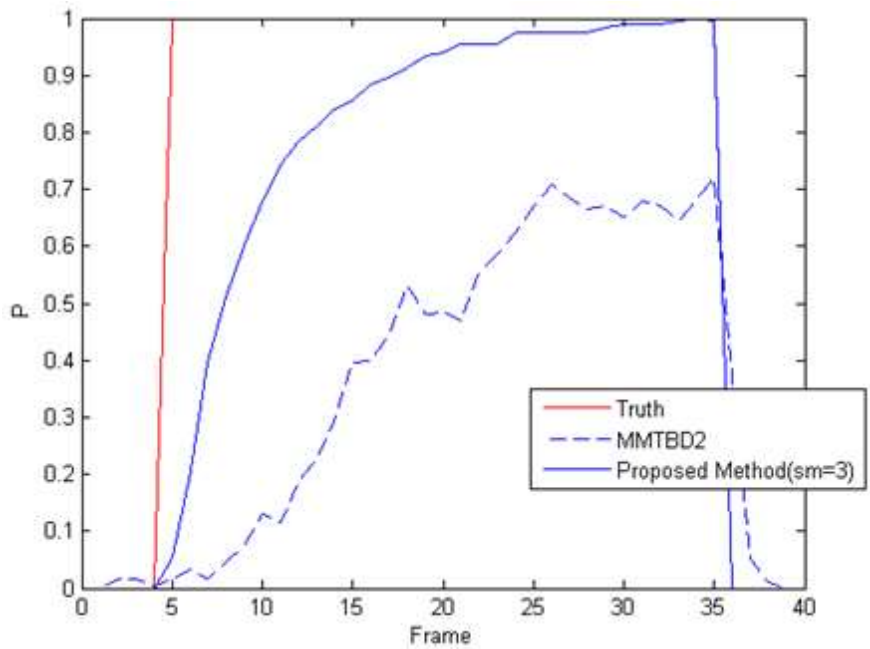

Fig 24: Detection probability (dash line: MMTBD2, solid line: SDT)

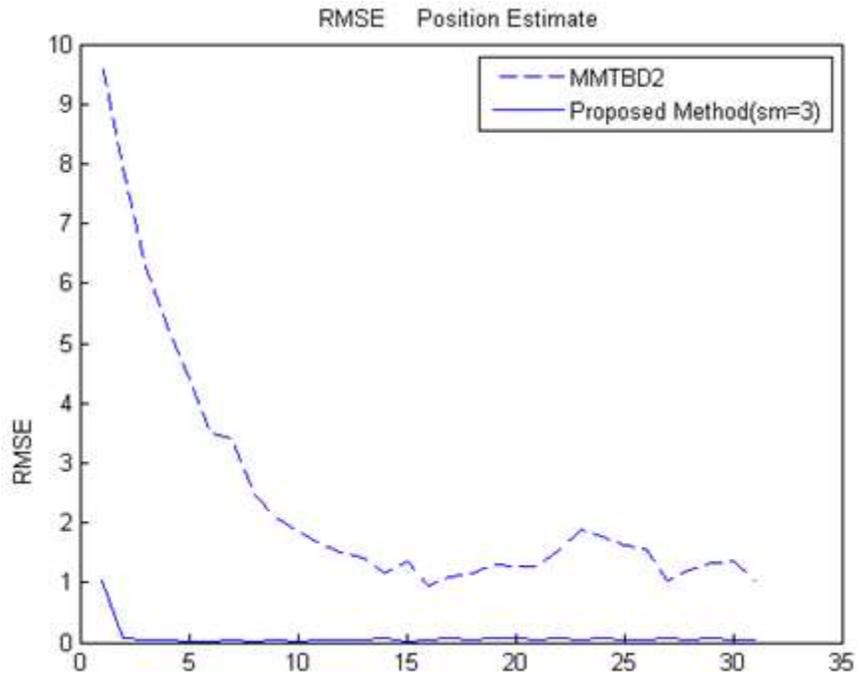

Fig 25: RMSE of position estimate (dash line: MMTBD2, solid line: SDT)

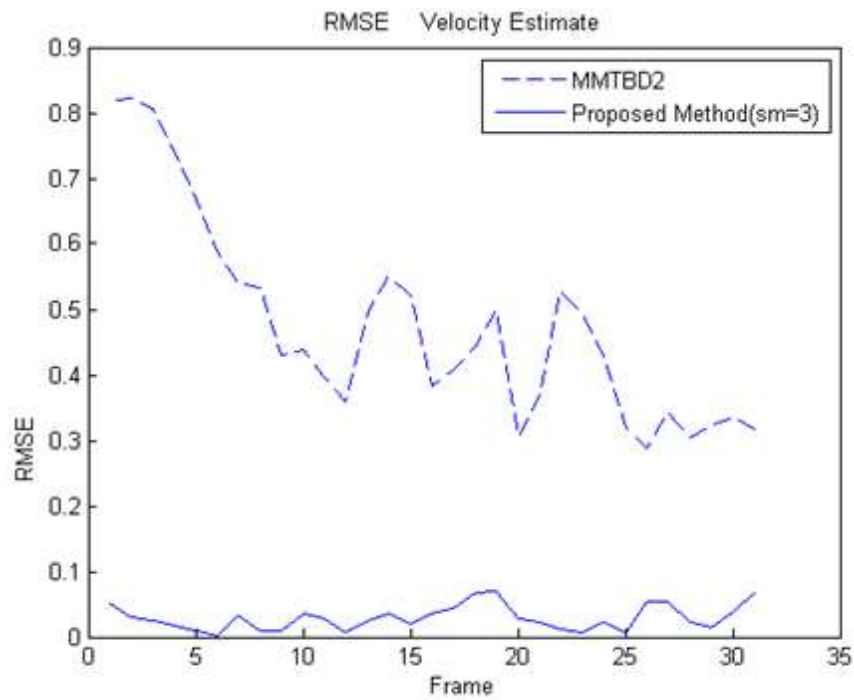

Fig 26: RMSE of velocity estimate (dash line: MMTBD2, solid line: SDT)

**Case III: SNR= 9dB**

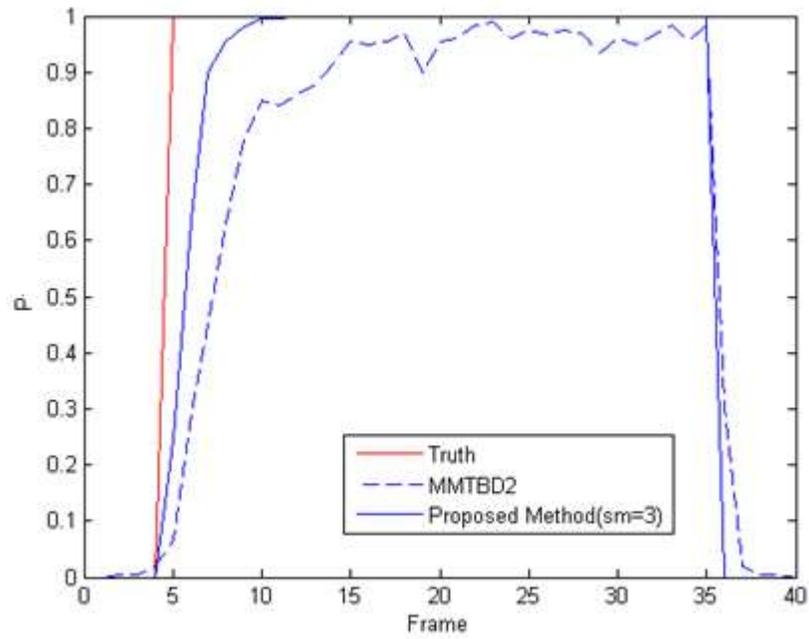

Fig 27: Detection probability (dash line: MMTBD2, solid line: SDT)

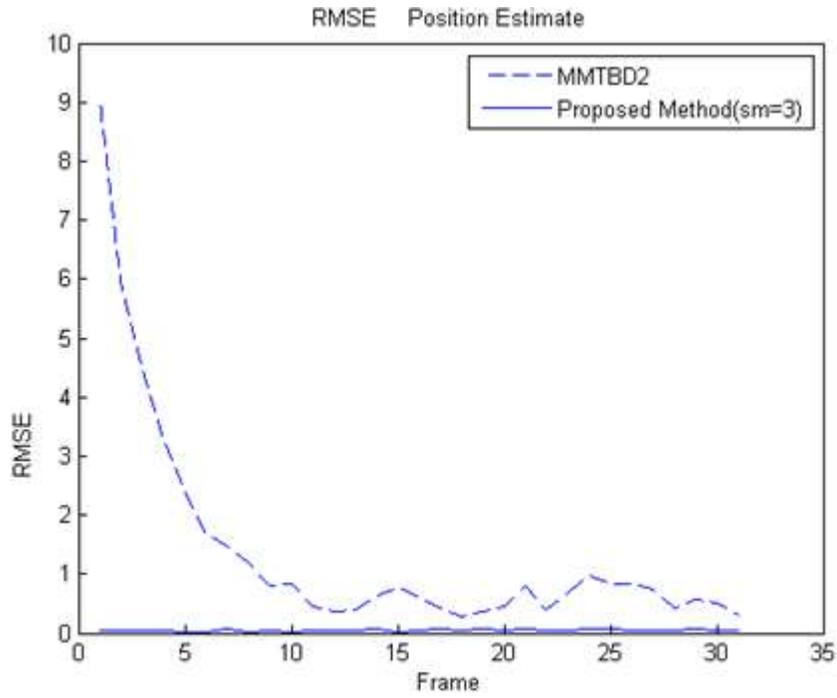

Fig 28: RMSE of position estimate (dash line: MMTBD2, solid line: SDT)

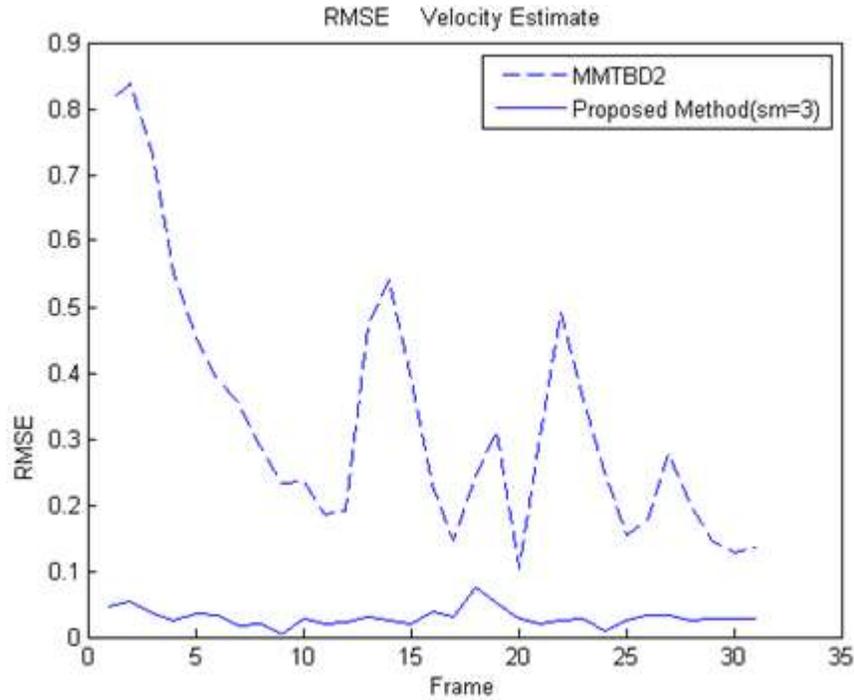

Fig 29: RMSE of velocity estimate (dash line: MMTBD2, solid line: SDT)

**Example 5:** SDT and MMTBD2 are compared in a high-speed high maneuver scenario.

- There exists an object from frame 5 to frame 35.
- SNR=3dB.
- Probability of detection during the whole 40 frames and RMSE of position and velocity estimates (when an object is detected) are plotted.
- For simplicity, SDT assumes at most one cell movement between two consecutive frames. The consequence of violation of this assumption in a high-speed scenario is studied. Also, a high maneuver is considered to compare the two approaches. The object has a high speed (moving more than one cell in two consecutive frames) once it appears at frame 5 until around frame 16. Then, it slows down and makes a sharp turn. Then, it again speeds up and moves more than one cell between two consecutive frames from around frame 29 to frame 35 when it disappears. This behavior can be seen from the trajectory of the object (Fig 30).
- SDT does not work well when the object moves more than one cell between two consecutive frames. That is because its assumption does not hold. This problem is resolved if we relax the maximum one-cell movement assumption in SDT and allow two (or more) cell movement in two consecutive frames. That increases the computational cost.
25

- SDT outperforms MMTBD2 during those frames that the object makes a sharp turn and moves not more than one cell in two consecutive frames (from frame 17 to 28).

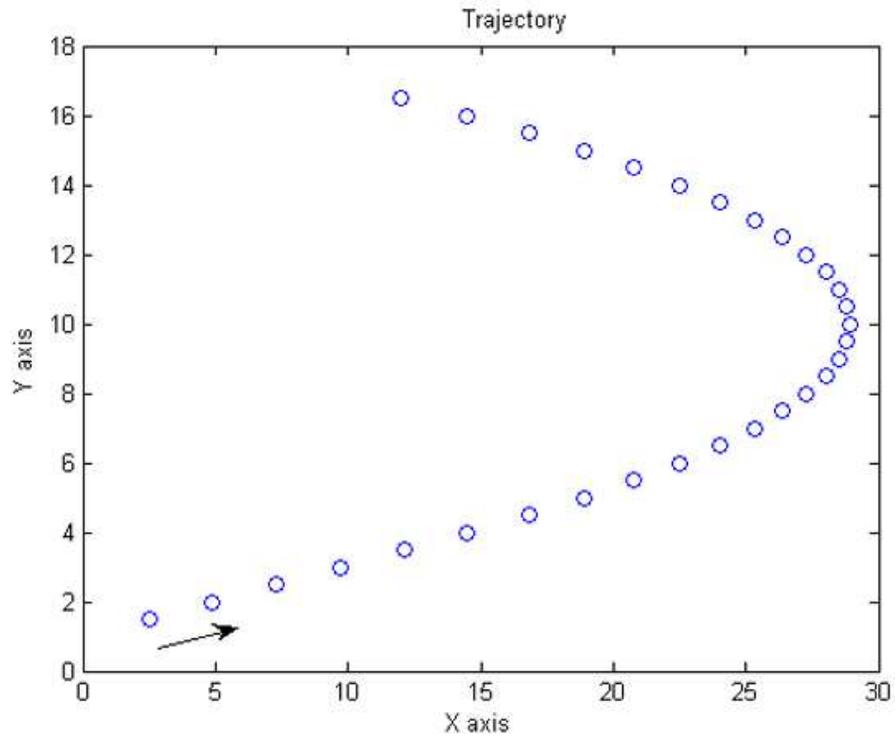

Fig 30: Trajectory

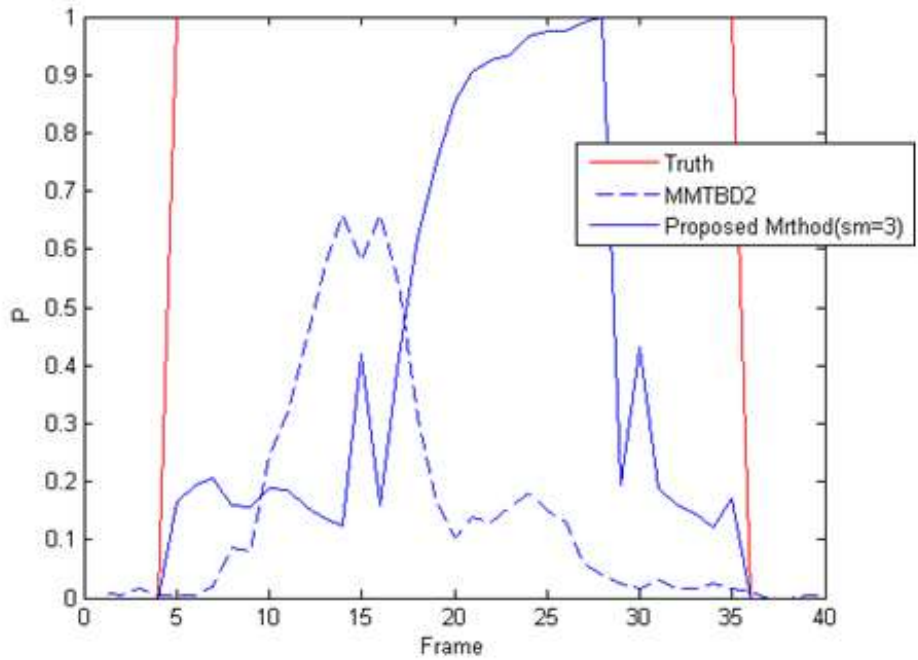

Fig 31: Detection probability

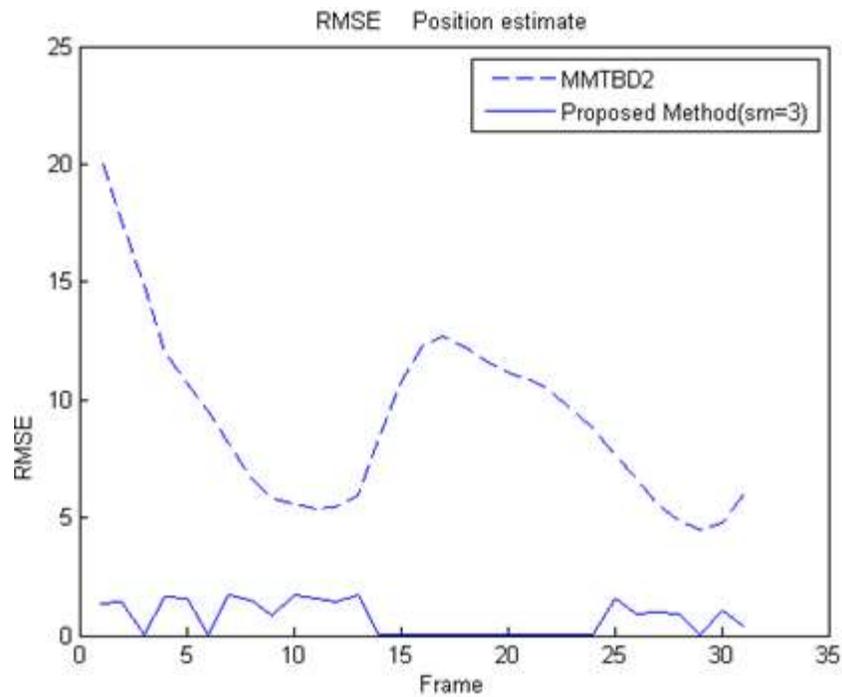

Fig 32: RMSE of position estimate

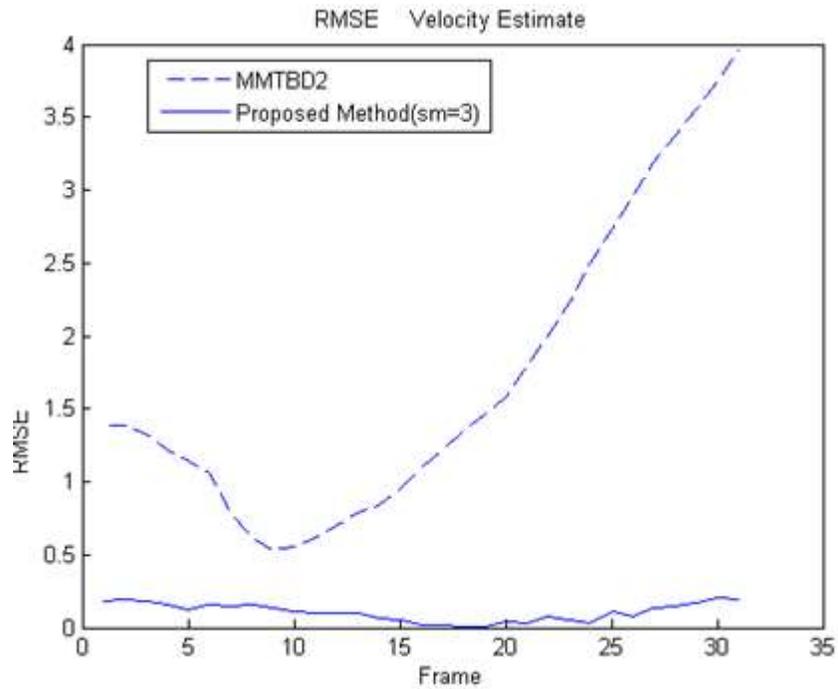

Fig 33: RMSE of velocity estimate

**Example 6:** SDT is evaluated for very low SNRs.

- There exists an object from frame 5 to frame 35.
- SNRs 1dB, 2dB, and 3dB.
- Probability of detection during the whole 40 frames and RMSE of position and velocity estimates (when an object is detected) are plotted.
- SDT can even handle very low SNR objects.

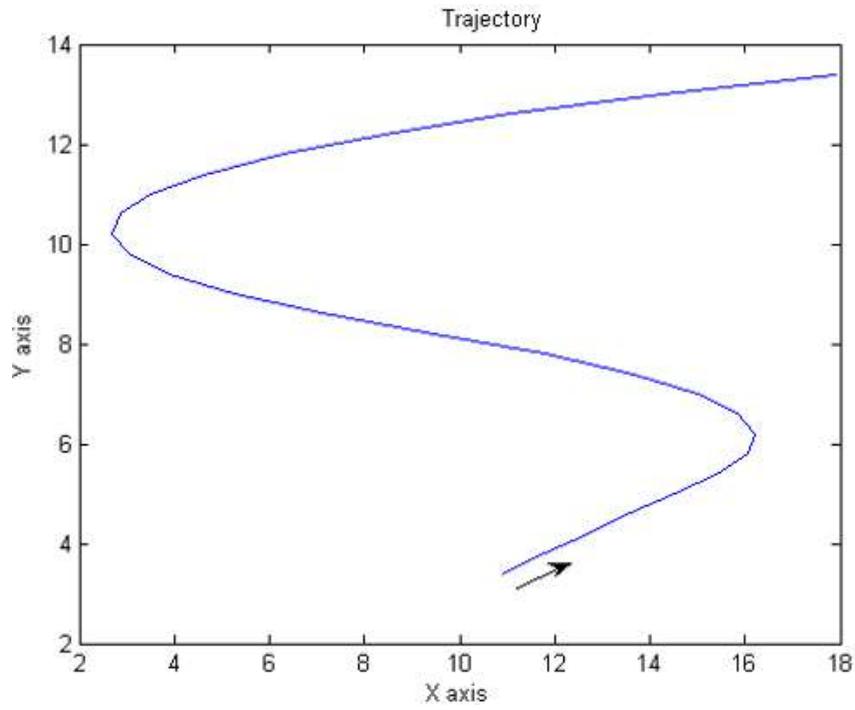

Fig 34: Trajectory

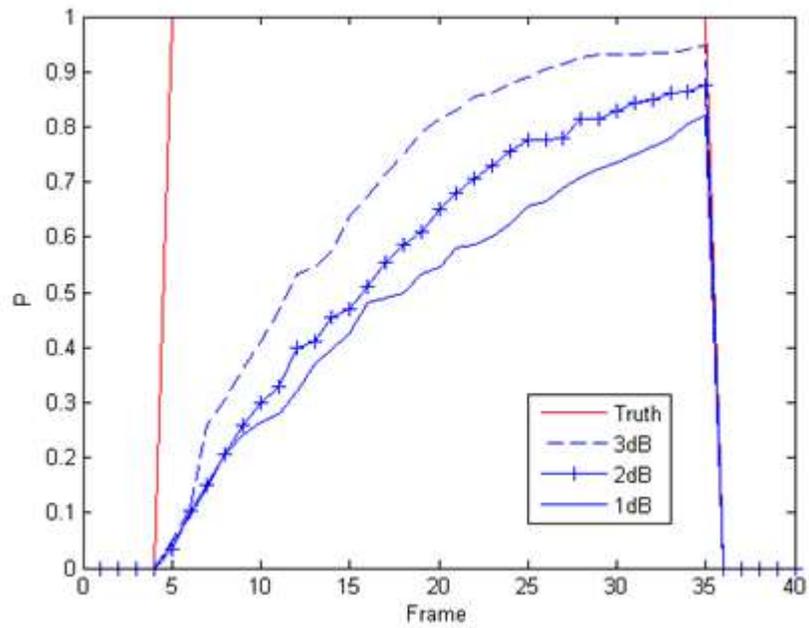

Fig 35: Detection probability (SDT for different SNRs)

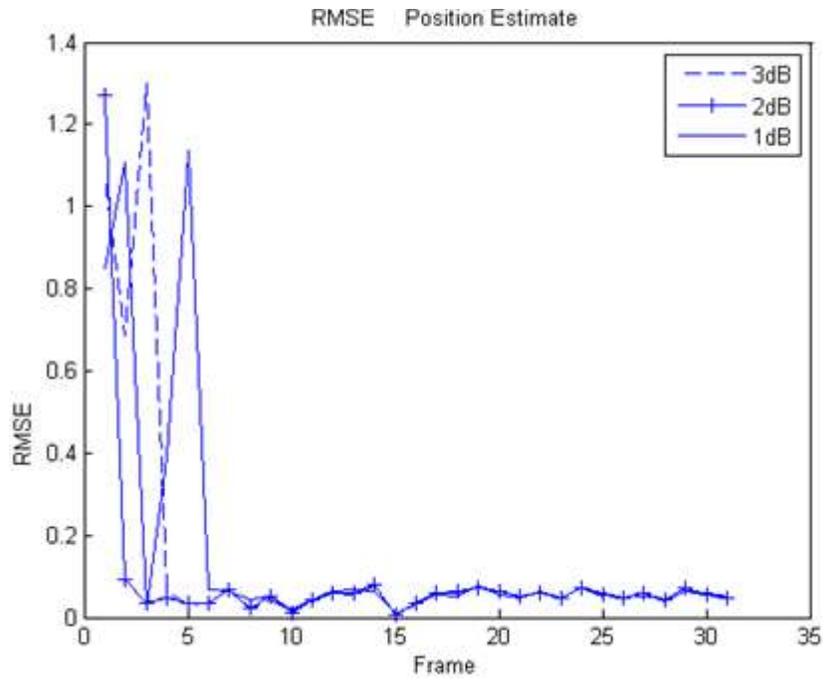

Fig 36: RMSE of position estimate

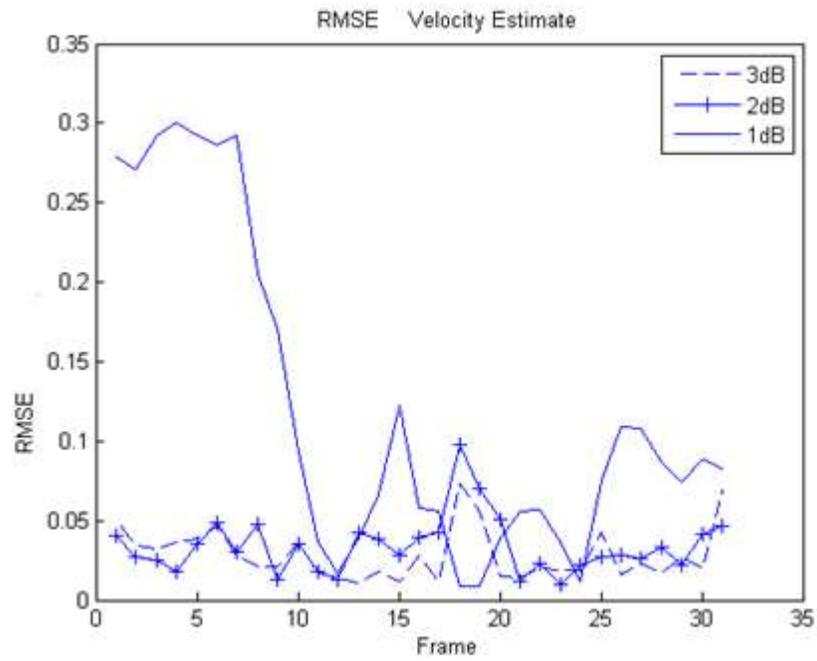

Fig 37: RMSE of velocity estimate

# Conclusions and Future Work Directions

## Conclusions

A sequential detection and tracking (SDT) approach for very low SNR objects has been proposed and its performance has been evaluated and compared with MMTBD approaches. It has been shown that the proposed SDT outperforms MMTBDs and demonstrated that it can even detect and track very low SNR objects. The SDT accumulates signals over consecutive frames until the SNR is high enough to detect even a very low SNR object with a high probability of detection.

## Future Work Directions

Incorporating more information about objects (e.g., their destination and waypoint) is helpful in their detection and tracking. A theoretical foundation of conditionally Markov (CM) sequences was presented in [27]-[35] and their application to trajectory modeling with fixed/moving destination and waypoint information were discussed in [36]-[40]. In the future, CM models can be used in the proposed SDT framework to model and incorporate fixed/moving destination and waypoint information and enhance the detection and tracking performance.

A joint Bayesian decision and estimation risk [41] has been used in some decision-estimation problems [41]-[43]. Detection and tracking of a low SNR object is a decision-estimation problem and joint Bayesian decision and estimation risk can provide a solution for that. In the future, such a solution can be derived and compared with the one proposed in this paper.